\documentclass{aastex631}

\begin{document}
\title{Novae: An Important Source of Lithium in the Galaxy
\footnote{Released on March, 1st, 2023}}

\author[0000-0003-2847-9650]{Jun Gao}
\affil{School of Physical Science and Technology, Xinjiang University, Urumqi, 830046, China}
\affil{Xinjiang Observatory, the Chinese Academy of Sciences, Urumqi, 830011, China}
\author{Chunhua Zhu}
\affil{School of Physical Science and Technology, Xinjiang University, Urumqi, 830046, China}
\author{Guoliang L\"{u}}
\affil{Xinjiang Observatory, the Chinese Academy of Sciences, Urumqi, 830011, China}
\affil{School of Physical Science and Technology,
Xinjiang University, Urumqi, 830046, China}
\author{Jinlong Yu}
\affil{College of Mechanical and Electronic Engineering, Tarim University, Alar 843300, PR China}
\author{Lin Li}
\affil{School of Physical Science and Technology, Xinjiang University, Urumqi, 830046, China}
\author{Helei Liu}
\affil{School of Physical Science and Technology, Xinjiang University, Urumqi, 830046, China}
\author{Sufen Guo}
\affil{School of Physical Science and Technology, Xinjiang University, Urumqi, 830046, China}

\correspondingauthor{Chunhua Zhu, Guoliang L\"{u}}
\email{chunhuazhu@sina.cn, guolianglv@xao.ac.cn}

\begin{abstract}
The source of the Galactic Lithium (Li) has long been a puzzle.
With the discovery of Li in novae, extensive research has been conducted.
However, there still exists a significant disparity between the observed abundance of lithium in novae and the existing theoretical predictions.
Using the Modules for Experiments in Stellar Astrophysics (MESA), we simulate the evolution of nova with element diffusion and appropriately increased the amount of $^3$He in the mixtures. Element diffusion enhances the transport efficiency between the nuclear reaction zone and the convective region on the surface of the white dwarf during nova eruptions, which results in more $^7$Be to be transmitted to the white dwarf surface and ultimately ejected.
Compared to the previous predictions, the abundance of $^7$Be in novae simulated in our model significantly increases. 
And the result is able to explain almost all observed novae.
Using the method of population synthesis, we calculate Li yield in the Galaxy. We find that the Galactic occurrence rate of nova is about 130\,yr$^{-1}$, and about 110\,M$_{\odot}$ Li produced by nova eruption is ejected into the interstellar medium (ISM).
About 73\% of Li in the Galactic ISM originates from novae, and approximately 15\%-20\% of the entire Galaxy.
It means that novae are the important source of Li in the Galactic.
\end{abstract}


\section{Introduction} \label{sec:intro}
Lithium (Li) is an extremely fragile element that is destroyed in the hydrogen\,(H)-capture reaction at temperatures as low as 2\,$\times$\,$10^{6}$K.
Its dominant isotope, $^7$Li, is the decay product of $^7$Be with a half-life of 53.3 days.
In the process of H burning, $^7$Be rapidly decays to $^7$Li through electron capture (pp\,II) after being formed via the proton-proton (pp) chain. 
$^7$Li itself is also destroyed through proton capture, resulting in very little lithium surviving in the H burning process.
This makes it almost impossible to accurately calculate the abundance of Li in most stars during their formation.
The production of primordial $^7$Li is a sensitive function of the baryon-to-photon ratio and can be estimated within the framework of standard primordial nucleosynthesis as long as the baryon density is obtained from the initial deuterium abundance or fluctuations of the cosmic microwave background (CMB) \citep{2003WMAP,2013WMAP,2020Planck}.
The expected primordial value is A(Li)$\approx$2.72\,dex \citep{1998Zyczkowski,2014Coc}, which is 3\,-\,4 times higher than the measured values in halo dwarf stars \citep{1982Spite,2020Planck}.
This difference is often referred to as the \textit{Cosmological lithium problem} \citep{2014Fields}.

Since the measured Li abundance in meteorites that preserves the protosolar in the interstellar medium (ISM) is A(Li)$\approx$3.3 dex \citep{2009Asplund,2009Lodders}, there is need for a galactic source to explain the increase from the initial value of 2.72\,dex.
This identification of the source is commonly known as the \textit{Galactic lithium problem}.
The ISM is the material that fills the space between the stars within a galaxy. It consists primarily of gas (atomic, molecular, and ionized) and dust. 
Planetary nebulae, supernova explosions, and novae explosions all inject a large amount of chemical elements into ISM.
The ISM is crucial for the formation and evolution of stars, and it plays a key role in the chemical enrichment and energy balance of galaxies \citep{2011Draine}.
Various nucleosynthesis processes and sources have been proposed so far.
One confirmed source of Li is spallation and fusion processes of galactic cosmic rays (GCRs) in the ISM.
The integrated spallation process is estimated to contribute about 10\%$\sim$20\% to the measured $^7$Li in the entire Galactic lifetime \citep{1993Prantzos,2001Romano,2012Prantzos}.
Other proposed sources include spallation processes in the flares of low-mass active stars, red giants (RGs), asymptotic giant branch (AGB) stars and neutrino-induced nucleosynthesis during a type II supernova \citep{1978Starrfield,1982Spite,1989Smith,1990Smith,1996Hernanz,1999Romano,2001Travaglio,2002Alib,2012Prantzos,2016Tajitsu,2016Banerjee,2016Pignatari,2017Rukeya}.
Although Li abundance is indeed enhanced in some studies, but their contribution to the entire Galaxy is too small.
Therefore, other sources are still needed for the Galaxy to reach its current value.
Subsequently, \cite{2001Romano} and \cite{2017Prantzos} found that low-mass giants are the best candidates for reproducing the late rise of the Li metallicity plateau.
However, high A(Li) cannot be sustained for a long period due to convective activity in these stars, and the low percentage of Li-rich giants also indicates this \citep{2016Casey,2018Yan,2022Gao}.
Therefore, their contribution to the enrichment of Li in the Galactic ISM remains quite uncertain.

For many years, classical novae have been proposed as feasible sites for Li production \citep{1975Arnould,1978Starrfield}.
The novae eruption is the result of unstable hydrogen-burning on the CO or ONeMg white dwarfs (WDs) surfaces, which accrete hydrogen-rich material from their main sequence (MS) or red giant (RG) phase companions in low-mass, close binary systems \citep{2009Starrfield,2016Jose,2016Starrfield,2017Rukeya}.
When the companion star fills its Roche Lobe, the hydrogen-rich material flows through the inner Lagrange point to the WD and accretes onto its surface.
This material accumulates and is compressed until thermonuclear runaways (TNR) is triggered, resulting in mass ejection that ultimately pollutes the interstellar environment.
Both theoretical and observational evidence confirms that novae contribute many nucleosynthetic isotopes to the Galactic ISM \citep{1998Jos,1998Starrfield}.
For example, the elements $^7$Li and $^7$Be.

$^7$Be is generated through the $^3$He($\alpha$,$\gamma$)$^7$Be reaction at a temperature of 150 million K \citep{1996Hernanz}.
To avoid destruction, $^7$Be needs to be transported to cooler regions quickly through convection, as described in the Cameron–Fowler\,(CF) mechanism \citep{1971Cameron}.
When these cooler regions are subsequently ejected, the absorption lines of $^7$Be during nova outbursts can be observed \citep{1998Jos}, eventually decaying into $^7$Li.
This view was later confirmed by observations by \cite{2015Izzo} who detected a potentially detectable $^7$Li I $\lambda$6708 Å absorption line in the spectrum of the Nova Centauri 2013 (V1369 Cen), providing observational evidence for the presence of Li in nova explosions that had been predicted since the mid-1970s but only recently discovered.
Subsequently, the observation studies continuously detected the progenitor nucleus $^7$Be or $^7$Li in the prominent post-outburst spectra of classical novae \citep{2015Tajitsu,2016Tajitsu,2016Molaro,2018Izzo,2018Selvelli,2020Molaro,2022Molaro,2023Molaro}.

For a long time, people have continuously used theoretical models to calculate the accurate Li yield from nova explosions.
\cite{1996Hernanz} and \cite{1998Jos} calculated the theoretical values for different WD masses and mixing ratios of ejected material in their nova models.
Then \cite{2017Rukeya} expanded on the nova grid model and computed more detailed mass and Li yields of the ejecta.
They concluded that the Li produced in novae accounts for 10\% of the total Galactic ISM ($\sim$\,150\,M$_{\odot}$) production.
However, \cite{starrfield2024} predicted that the abundance of Li ejected from CO novae explosions is A(Li)\,$\approx$\,6, but most observed novae showed A(Li)\,$\textgreater$\,7 (i.e.\,A(Li)=Log(N($^7$Li)/N(H)+12)).
The actual abundance of ejected Li is one order of magnitude higher than the theoretical predictions \citep{2015Izzo,2016Molaro,2018Izzo,2021Arai,2022Molaro}.

This suggests that traditional nova models may have deficiencies in their physical mechanisms, and the Galactic lithium problem unresolved.
Subsequently, researchers have attempted different calculation methods by adjusting the mass distribution of binary systems, eruption time intervals, criteria for eruptive episodes, metallicity, and accretion material mixing ratios to compute Li yields under different scenarios \citep{1998Jos,2009Starrfield,2019Cescutti,2020Jos,2020Starrfield}. 
\cite{2022Kemp} demonstrated that using the yield values given in \cite{2023Molaro}, then novae can definitely be the main factories of lithium in the Galaxy while using the theoretical values, they could not.
Therefore, finding a reasonable nova model and Li production mechanism that aligns with the observations and theoretical values is of great significance in addressing the Galactic lithium problem and even the Cosmological lithium problem.
It is well known that the chemical composition of the ejecta in nova explosions depends on the surface chemical element composition of the WDs at the moment of the eruption.
Several factors can influence the chemical composition of the WD's surface, such as metallicity, accretion efficiency, and element diffusion \citep{1980Kippenhahn,1992Dupuis,2021Zhu}.
The effects of nuclear reaction rate, accretion efficiency and metallicity on the surface material of WDs have been explored by \cite{2016Starrfield,2020Starrfield} and \cite{2022Kemp,2022Kemp1}.
However, element diffusion has been rarely considered.
Element diffusion is a dynamic process that alters the distribution of chemical elements within a star.
It primarily results from the combined effects of pressure, temperature, and material concentration.
\cite{1985Kovetz} investigated that effective diffusion between the accreted layer and the inner regions for accreting WD, and they found that the diffusion can lead to enrichment of CNO (and other metals) in the ejected envelop.
However, CNO enhancements are possible in CO novae only in the absence of the He buffer \citep{1991Iben}. The presence of helium layer would prevent the diffusion of CNO and metals (such as $^3$He), although the average abundance of helium is almost the same as that of solar \citep{1998Gehrz,2005Yaron}. 
Therefore, the $^3$He inside the WD cannot be effectively brought to the envelope where TNR occurs by element diffusion, and there are other sources of $^3$He in the envelope \citep{1991Iben}.
In our model, $^3$He comes from the accreted material from the donor star but not the from the interior of WD. The role of element diffusion in our model is that $^7$Be produced by $^3$He($\alpha$,$\gamma$)$^7$Be among TNR can be taken more efficiently to WD surface, and then can ejected more easily.
Therefore, a large amount of $^7$Be can be detected in the ejecta of the nova, ultimately decaying into $^7$Li.

In this paper, we assume the novae as main sources of Li.
In particular, Section 2 describes the details of the nova model.
Section 3 presents the $^7$Li yields predicted by our models and the comparative analysis of the observation and theoretical results.
The summary is given in Section 4.

\section{Models} \label{sec:style}
Our nova model has been using version-12778 of the MESA stellar evolution code, which constructs CO and ONeMg nova models based on its \textit{white dwarf} and \textit{nova} modules \citep{2011Paxton,2013Paxton,2015Paxton,2018Paxton,2019Paxton}.
The advantage of MESA is its ability to calculate nova outbursts by dividing the outer shell mass of WD into approximately 1000 or more cells, resulting in smaller errors.
At each cell, such as temperature, density and isotopic composition are calculated.
In MESA, when the luminosity of a star ($L$) exceeds the super-Eddington luminosity ($L_{\rm Edd}$), it will trigger mass loss. 
The mass loss rate is
\begin{equation}
\dot{M} = -2\eta_{\rm Edd}\frac{(L-L_{\rm Edd})}{v_{\rm esc}^2}\label{gshi}
\end{equation}
where $v_{\rm esc}=\sqrt{2GM/R}$, $L_{\rm Edd}=(4\pi GcM)/\kappa$. $M$ and $R$ are the mass and radius, while $\kappa$ is the Rosseland mean opacity at the WD’s surface. The scaling factor is taken $\eta_{\rm Edd}=1$ \citep{2013Denissenkov}. The cells will be ejected when $L>L_{\rm Edd}$.
In this paper, it is set that the nova ejection begins when the total WD luminosity ($L$) is greater than $10^{4}$ times the solar luminosity ($\rm L_{\odot}$), and ends when less than $10^{3}$\,$\rm L_{\odot}$.
Therefore, the MESA code can be used to simulate multi-cycle nova and construct a large-scale nova model grid from the accretion phase to expansion, explosion and ejection phases \citep{2011Paxton,2013Paxton,2015Paxton}.

For CO WD models, the nuclear network selects \textit{pp\_and\_cno\_extras\_net}, while the ONeMg WD models uses \textit{h\_burn\_net}.
These nuclear networks include the CNO burning cycle and the proton-proton reaction chains (pp chain), the latter of which include $^3$He($\alpha,\gamma$)$^7$Be, $^7$Be($e^-,\nu$)$^7$Li, $^8$B($\gamma,p$)$^7$Be, $^7$Li($p,\alpha$)$^4$He, and $^8$B($e^+,\nu$)$^8$Be$^*$(2$\alpha$), which are sufficient nuclear synthesis to treat $^7$Be and $^7$Li nucleosynthesis \citep{1978Starrfield}.
It is worth noting that the degree of mixing also affects the mass and isotopic abundance of the ejected material \citep{1996Hernanz}.
\cite{1998Jos}, \cite{2014Denissenkov} and \cite{2017Rukeya} have used the NOVA and MESA codes to calculate the nova eruption model, assuming that the degree of mixing could be 25\%-50\%.
That is, 25\% of WD material and 75\% of solar material, or 50\% of WD material and 50\% of solar material \citep{2009Lodders}.
We adopted the latter, which is more widely used.
However, the $^7$Be simulated by the previous models did not explain the observed values.
Subsequent studies suggested that there is a corresponding relationship between the $^7$Be and $^3$He \citep{1993Boffin,1996Hernanz,2020Molaro,2021Denissenkov}.

The importance of $^3$He for novae was first studied by \cite{1951Schatzman} in the context of a theory of novae powered by thermonuclear detonations. 
The initial $^3$He abundance for the accreted matter, could potentially come from the donor star \citep{1980Shara,2004Townsley}.
As the donor star are ascending the red giant branch the convection dredges up $^3$He enriched material to the surface which is later expelled into the ISM by wind or is accreted by WDs in nova systems \citep{2009Shen,2021Denissenkov}.
\cite{1982Dantona} and \cite{1984Iben} found that the mass fraction of accreted $X_{0}$($^3$He) can reach values as high as $4 \times 10^{-3}$ during the evolution of systems with low-mass donors.
Furthermore, some observational studies have shown the existence of relatively high levels of $^3$He/H in certain planetary nebulae \citep{1992Rood,2018Balser}.
If $^3$He abundance of the donor's envelope is much higher than that in the Sun, this would almost align the model predictions of $^7$Be production with observations \citep{2020Molaro}.
However, observations of the ISM indicate that the abundance of $^3$He in the Galaxy cannot be too high \citep{1996Dearborn,2003Romano}.
\cite{2021Denissenkov} demonstrated that an excess of $^3$He in accreted material would actually reduce the production of $^7$Be.
Their work showed that when $X_{0}$($^3$He) exceeds $3\times10^{4}$, it could lead to the early onset of the TNR, result in a reduction in peak temperature and accreted mass, and thereby suppress the production of $^7$Be.
While this level of $^3$He-induced $^7$Be abundance increase would be slightly elevated, most of it would still be consumed at high temperatures and not transported to the WD surface to generate sufficient $^7$Li, thus not closing the gap with observations.
Therefore, an effective transport mechanism is required to safely transport $^7$Be to the WD surface and allow it to survive.
Element diffusion provides an effective pathway for this.

Elemental diffusion in stellar interior is mainly driven by a combination of pressure gradients (or gravity), temperature gradients,  compositional gradients, and radiation pressures.
By solving the Burgers equation \citep{1969Burgers}, \cite{1994Thoul} proposed a general method of arranging the entire system of equations into a single matrix equation, so that the relative concentrations of various species have no approximate values and the number of elements considered is not limited.
Therefore, this method is applicable to various astrophysical problems.
In MESA, the method of \cite{1994Thoul} is used to calculate the diffusion of chemical elements within stars \citep{2015Paxton,2018Paxton}.
This diffusion effect can bring internal elements to the surface.
We speculate that when WD accretes enough hydrogen-rich material from companion star to reach the critical point, TNR occurs. 
The thermal instability caused by the thermonuclear runaway leads to the appearance of a convective zone, which extends from approximately the middle of the combustion shell to the surrounding non combustion helium layer \citep{1971Cameron}. 
The convective zone can bring $^7$Be from the stellar interior to the surface, which is the CF mechanism. 
Element diffusion improves the mixing efficiency of the entire convection zone, bringing more $^7$Be to the surface.
When a nova explosion occurs, a large amount of $^7$Be on the surface is ejected and decays into $^7$Li in the ISM through a short half-life.
This work utilize MESA to construct a large-scale nova grid by setting different basic parameters and calculated the $^7$Li production of the nova.

MESA is one-dimensional code, and only simulate spherically symmetric nova outburst. 
However, based on the images of NOVA V5668 SAG combining the optical and radio observations from the Hubble Space Telescope and the Very Large Array telescope, \cite{2019Mukai} suggested that the geometry distribution of nova ejecta is not spherical symmetry but is the shape of an equatorial torus. 
It indicates that ejecta from novae has very complicated structure. 
It is unclear which physical process led to such an unsymmetric structure. 
Up until now, there have been few instances of 2D or 3D simulations depicting nova outbursts.
Therefore, it should be noted that the MESA nova model also has its limitations.

\section{result} \label{sec:floats}
According to \cite{2005Yaron}, the WD mass\,($M_{\rm WD}$), accretion rate\,($\dot{M}$), WD core temperature\,($T_{\rm c}$), and composition of the accreted material constitute the four basic input parameters that determine the nova eruption \citep{2016Starrfield,2016Jose}.
\cite{2005Yaron} constrained the parameters of the nova eruption and set a three-dimensional restricted space for the nova eruption conditions.
Our model set the $T_{\rm c}$ at 3$\times$10$^7$K and selected CO WD model mass: 0.5\,$\rm M_{\odot}$, 0.6\,$\rm M_{\odot}$, 0.7\,$\rm M_{\odot}$, 0.8\,$\rm M_{\odot}$, 0.9\,$\rm M_{\odot}$, 1.0\,$\rm M_{\odot}$, 1.1\,$\rm M_{\odot}$, 1.2\,$\rm M_{\odot}$; and ONeMg WD model mass: 1.0\,$\rm M_{\odot}$, 1.1\,$\rm M_{\odot}$, 1.2\,$\rm M_{\odot}$, 1.3\,$\rm M_{\odot}$; with accretion rates\,($\dot{M}$) of 1\,$\times$\,$10^{-11}$\,$\rm M_{\odot}$\,yr$^{-1}$, 1\,$\times$\,$10^{-10}$\,$\rm M_{\odot}$\,yr$^{-1}$, 1\,$\times$\,$10^{-9}$\,$\rm M_{\odot}$\,yr$^{-1}$, 1\,$\times$\,$10^{-8}$\,$\rm M_{\odot}$\,yr$^{-1}$, 1\,$\times$\,$10^{-7}$\,$\rm M_{\odot}$\,yr$^{-1}$.
In addition to the above four parameters, it is expected that material transferred from the companion (solar-like) will mix with the outer layers of the WD.
This model adopts the widely accepted composition of 50\% WD material and 50\% solar material.
Solar component data is taken from \cite{2009Lodders}.
\cite{2020Molaro} argued that the variety of high $^7$Be/H abundances in nova could be originated in a higher than solar content of $^3$He in the donor star.
Observations show relatively high levels of $^3$He/H in some planetary nebulae, but observations of ISM indicate that the abundance of $^3$He in the Galaxy cannot be too high \citep{1992Rood,1996Dearborn,2003Romano,2018Balser}.
According to the $X_{0}$($^3$He) range: 2.96$\times$10$^{-5}$$\thicksim$2.96$\times$10$^{-3}$ used by \cite{2021Denissenkov}, we set $X_{0}$($^3$He) to be 4$\times$10$^4$.

\subsection{Novae models as stellar sources of Lithium}
\begin{figure}[ht!]
\plotone{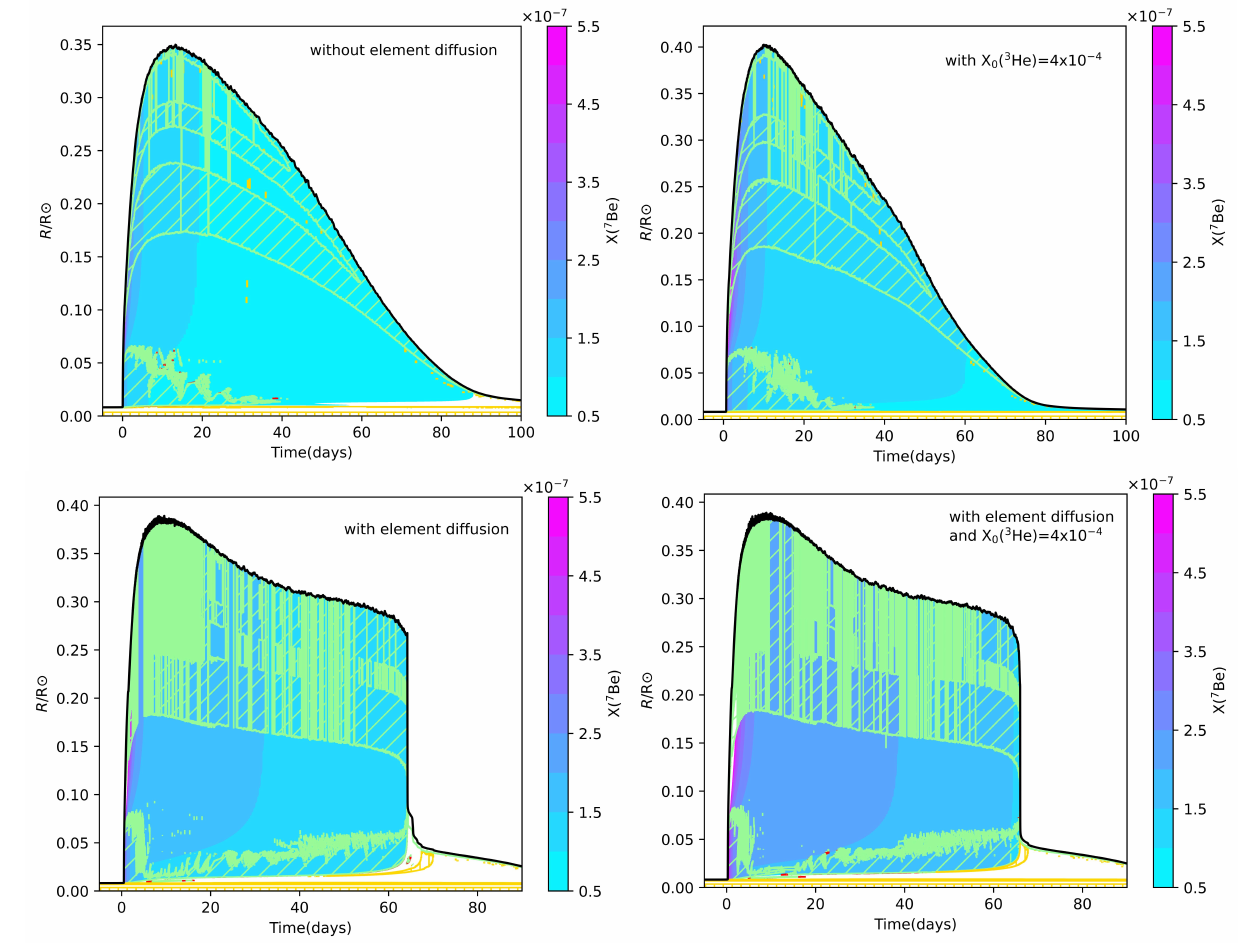}
\caption{Nova explosion process with $M_{\rm WD}$\,=\,1$\rm M_{\odot}$ and $\dot{M}$\,=\,1\,$\times$\,$10^{-10}$\,$\rm M_{\odot}$\,yr$^{-1}$.
Profiles of $^7$Be during the nova explosion show in panels.
The y axis represents the boundary radius of WD during explosion and the x axis represents the lifetime of the nova explosion.
Green shadow shows the convective zone of different models.
\label{fig:1}}
\end{figure}

It is well known that during the occurrence of TNR, there are intense nuclear reactions, primarily dominated by hydrogen burning due to the WD accreting hydrogen-rich material from the companion.
During hydrogen burning, a large amount of $^7$Be is produced through $^3$He($\alpha$,$\gamma$)$^7$Be, with some being convectively transported to the WD’s surface while the rest is depleted at high temperatures.
In our models with element diffusion, the transport efficiency within the WD is enhanced.
Figure \ref{fig:1} shows the evolution of elements during nova explosion.
The left two panels clearly display the effect of element diffusion. 
There are some blank discontinuous area on the surface convection zone (green shadow) of the model without element diffusion, indicating weak or even no mixing activity in these areas, as shown in the left-upper panel. 
In Figure \ref{fig:1}, the mixing area within the convection zone of the WD model with element diffusion becomes more continuous and dense, and the mixing activity becomes stronger, as shown in the left-lower panel.
In the nova model with element diffusion, a large amount of $^7$Be generated by nuclear reactions is effectively transported by this strong mixing effect to the low-temperature region of the surface, thereby avoiding depletion at high temperatures.
On the contrary, the model without element diffusion cannot send out more $^7$Be, which is consumed by high temperature.

Meanwhile, the improvement of $^3$He can generate more $^7$Be to contribute to the transport to the surface as shown by the right-top panel in Figure \ref{fig:1}.
The $^3$He increase greatly enhances the yield of $^7$Be in envelope, which is shown by pink color area. However, the increase of $^3$He causes the convective region to move inward, resulting in the failure to bring $^7$Be to a sufficiently low temperature surface.
This result is consistent with \cite{2021Denissenkov}, who stated that excessively high $^3$He actually reduces the $^7$Be of nova ejection.
However, combining the element diffusion and the increase of $^3$He, the large amount of $^7$Be produced via enhancing $^3$He can be efficiently transfered to WD surface, which are clearly shown by the right-lower panel. 
\begin{figure}[ht!]
\plotone{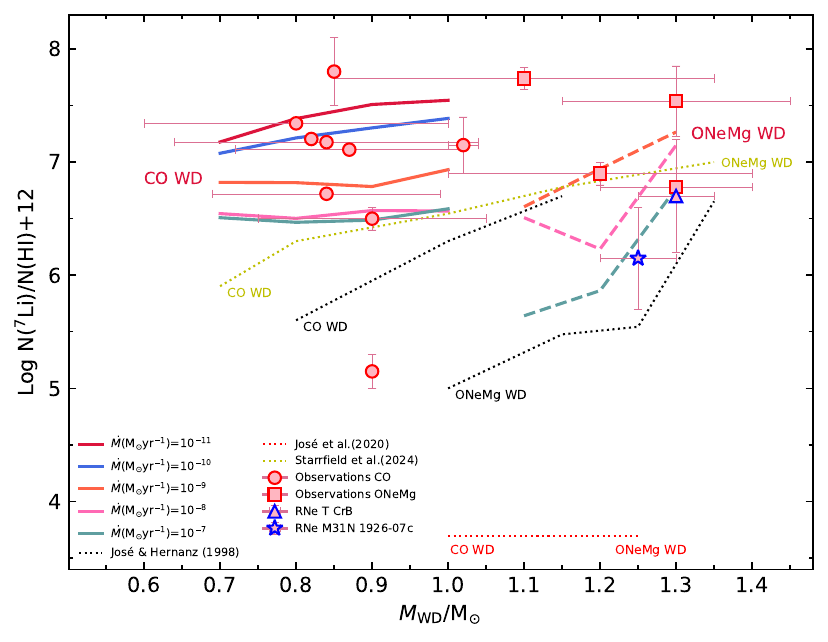}
\caption{Comparison with observed Li abundances and the numerical simulations along with WD masses.
Solid lines of different colors represent CO nova models with different accretion rates, while dashed lines represent ONeMg nova models.
Red bordered circles denote the observed abundances of CO WD novae, and the squares denote the ONeMg WD novae.
Since evaluating the quality of each WD is difficult, we simply plotted the error bars for each observed nova.
The black, yellow, and red dotted lines correspond to theoretical predictions; highest
values for each WD mass in \cite{1998Jos}, the “123-321 models” of \cite{2020Jos}, and 25\%–50\% mixing model of \cite{starrfield2024}, respectively.
The blue pentagram represents the prediction of nova M31N 1926-07c and the blue triangle represents the nova T CrB.
\label{fig:2}}
\end{figure}
We find that the abundance of $^7$Be in the model combining the element diffusion and the increase of $^3$He can be increased by about 4\,-\,10 times. 
It means that $^7$Be abundance calculated in this work is consistent with the observational values \citep{2015Izzo,2015Tajitsu,2020Molaro,2023Molaro}.

In order to compare our results with the observations, we selected 13 published observational samples of CO and ONeMg novae with relatively accurate measurements of parameters such as mass and Be abundance (9 CO novae and 4 ONeMg novae) (e.g. \cite{2015Tajitsu,2016Tajitsu,2016Molaro,2018Izzo,2018Selvelli,2020Molaro,2021Arai,2022Molaro,2023Molaro}).
We calculated the $^7$Be abundance ejected in a single nova eruption using the criteria that the eruption begins when the total WD luminosity ($L$) is greater than $10^{4}$\,$\rm L_{\odot}$ and ends when it is below $10^{3}$\,$\rm L_{\odot}$.

Figure \ref{fig:2} presents a comparison of the Be abundance produced by our simulated nova models with other models and matching to observations.
The solid lines in different colors represent the CO nova models in the mass range of $0.7\rm M_{\odot}-1.0\rm M_{\odot}$, with accretion rates from 1\,$\times$\,$10^{-7}$\,$\rm M_{\odot}$\,yr$^{-1}$ to 1\,$\times$\,$10^{-11}$\,$\rm M_{\odot}$\,yr$^{-1}$. The bold dashed lines depict the results of the ONeMg nova models in the mass range of $1.1\rm M_{\odot}-1.3\rm M_{\odot}$, with accretion rates from 1\,$\times$\,$10^{-7}$\,$\rm M_{\odot}$\,yr$^{-1}$ to 1\,$\times$\,$10^{-9}$\,$\rm M_{\odot}$\,yr$^{-1}$.
The results simulated by other literatures also are given, such as the black and red dotted lines, representing models constructed by \cite{1998Jos,2020Jos} with 50\% mixing and different WD masses, and the yellow dotted line representing models constructed by \cite{starrfield2024} with 25\% and 50\% mixing and different WD masses.
It is worth mentioning that our model provide $^7$Be yields for a range of accretion rates for each $M_{\rm WD}$, rather than sampling only at a fixed accretion rate of 2\,$\times$\,$10^{-10}$\,$\rm M_{\odot}$\,yr$^{-1}$ as in the other models mentioned above.
It is evident that the $^7$Be abundance produced by our model is significantly higher compared to other models and covers almost all CO and ONeMg nova observational samples.
In addition, we have made predictions for the two upcoming nova T CrB and M31N 1926-07c eruptions in 2024 and marked the expected abundance of ejected $^7$Be in Figure \ref{fig:2}.
This result indicates that element diffusion enhances the transport efficiency between the interior and surface convective region of the WD, transporting sufficient $^7$Be to the surface region.
Our results are consistent with observations and successfully explain the high $^7$Be abundance observed in the spectra after nova eruptions.
\begin{figure}[ht!]
\plotone{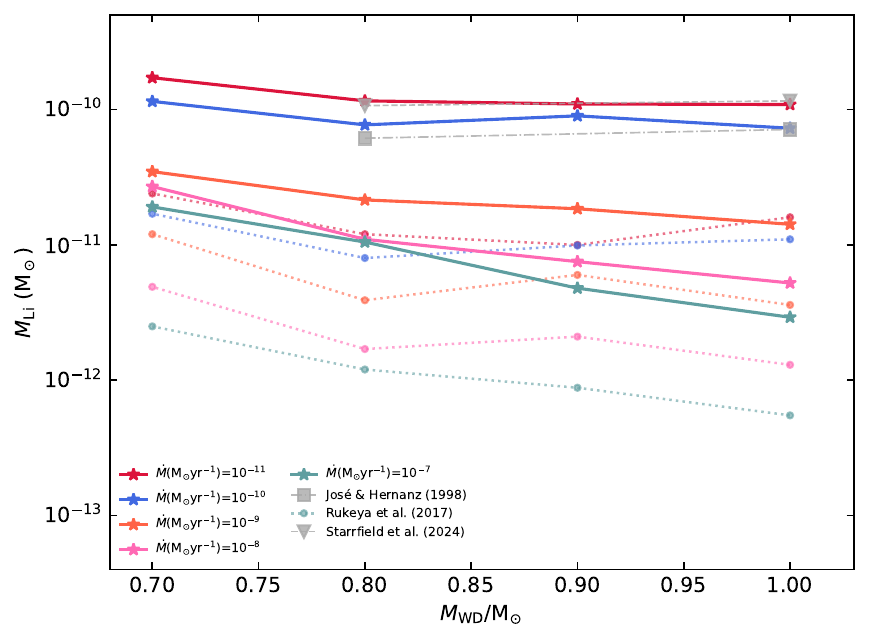}
\caption{$^7$Li yields vs. WD mass for all CO WDs in the grid.
\label{fig:3}}
\end{figure}

Due to the intense shock produced by nova eruption, the ash of TNR is partially ejected into the ISM \citep{2015Tajitsu,2015Izzo,2022Molaro,2023Molaro,starrfield2024}. 
The ejected $^7$Be during nova eruption decays into$^7$Li. 
This contributes to the total Li elements in the Galactic ISM.
Therefore, we consider the abundance of ejected $^7$Be to be $^7$Li.
In our simulations, Li mass produced by a nova outbursts is calculated by 
$M_{\rm Li}=\int \dot{M}X({\rm ^7Li}) {\rm d}t$, where $\dot{M}$ is the mass-loss rate of WD during nova outburst (See Eq.\ref{gshi}), $X(^7{\rm Li})$ is the mass fraction of $^7$Li in the nova ejecta.
Figure \ref{fig:2} shows $X(^7{\rm Li})$. 
Compared to the previous theoretical results in \cite{1998Jos}, \cite{2020Jos} and \cite{starrfield2024}, the Li abundances in our simulations are higher and are closer to the observed values. 
Figure \ref{fig:3} gives the Li yields for a nova outbursts. 
Our results are consistent with those of \cite{starrfield2024}. 
Based on Figures \ref{fig:2} and \ref{fig:3}, our model may underestimate the mass-loss rate $\dot{M}$, which is described by Eq.\ref{gshi}. 
Observations indicate that the range of ejection masses falls between 10$^{-6}$ to 10$^{-4}$\, M$_{\odot}$, and \cite{1998Jos} proposed an average ejection mass of $2\times10^{-5}$\, M$_{\odot}$ for novae. 
In our models, we calculate the range of ejection masses falls between $9.17\times10^{-7}$ to $5.93\times10^{-5}$\,M$_{\odot}$. 
This result is slightly smaller than the observed value. 
As mentioned by \cite{2021Denissenkov}, the increase of $^3$He in the accretion layer of our WD model can lead to an earlier TNR, a decrease in peak temperature and accreted mass, and ultimately result in a smaller ejection mass. 
By utilizing the Li and Be abundances and the ejected mass of WDs output by MESA, we can obtain the Li mass for each burst ejection.

The known novae are mainly concentrated in the range of WD mass $M_{\rm WD}$=0.8\,M$_{\odot}$\,-\,1.0M\,$_{\odot}$ and accretion rates around $1\times10^{-10}$\,$\rm M_{\odot}$\,yr$^{-1}$.
In this range, our estimating Li yield reaches to about $2\times10^{-10}$\,M$_{\odot}$. 
This yield is slightly lower than the observations of the latest recurrent novae RS Oph \citep{2015Izzo,2023Molaro}, but it has already reached the same order of magnitude.
It can be demonstrated that the nova model constructed in this paper can explain a large part of the observations in the Galaxy.

\subsection{Contribution of $^7$Li produced by novae to the Galactic} \label{sec:cite}
Theoretically, \cite{2017Rukeya} and \cite{starrfield2024} estimated the specific contribution of $^7$Li produced by novae to the Galactic, and they found that this contribution is approximately 10\%.
However novae have been proposed as the main factories of lithium in the Galaxy is something that was proposed in the 1970s by \cite{1975Arnould} and \cite{1978Starrfield}, and then demonstrated in \cite{2015Izzo}, \cite{2015Tajitsu} and \cite{2016Molaro,2023Molaro}, based on the detection of lithium and beryllium in a few novae.
As shown by Figure \ref{fig:3}, compared with the observations, the theoretical predictions underestimate Li yields produced by nova eruption.  

In order to estimated the contribution of $^7$Li produced in our nova models to the Galactic ISM, we use the method of binary population synthesis (BPS), which has been applied by our group \citep{2006Lv,2009Lv,2013Lv,2017Rukeya,2019Yu,2020Lv,2021Yu,2022Gao}.
BPS is a robust approach to evolve a large number of stars (including binaries) so that we can explain and predict the properties of a population of a type of stars \citep{2020Han}.
In the population synthesis method for binary systems, input parameters include the star formation rate (SFR), the initial mass function (IMF) of the primaries, the initial mass ratio distribution, the initial orbital distribution, the eccentricity distribution, and the
metallicity Z of binary systems.
Here, with help of the rapid binary evolution (BSE) code, originating from \cite{2002Hurley}, we can rapidly evolve a large-sample binaries into nova binaries, in which $M_{\rm WD}$ and the $\dot{M}$ are given.
\begin{figure}[ht!]
\plotone{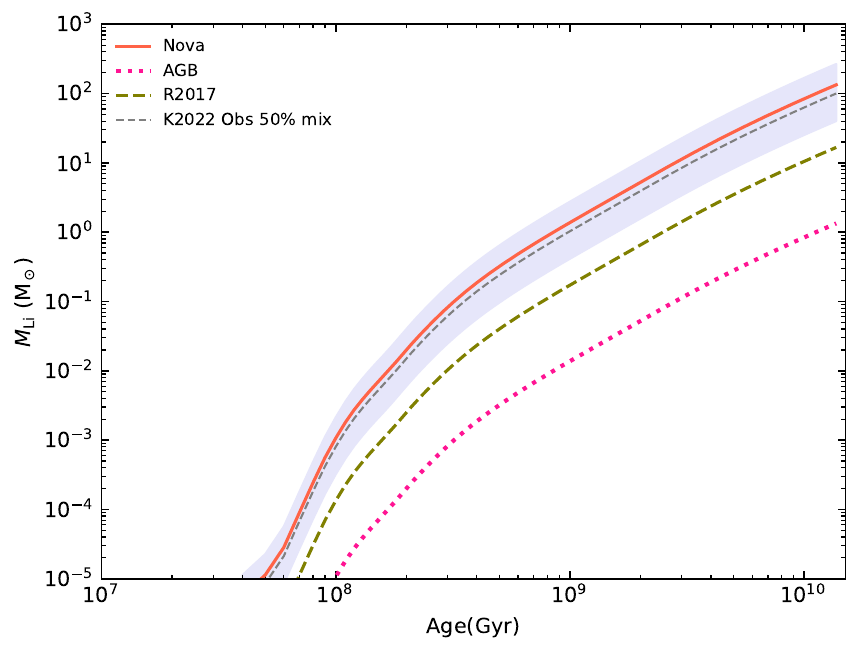}
\caption{Li mass contributions from different sources as a function of Galactic
age.
The orange line represents our work, while the green and gray dashed lines represent the model results for \cite{2017Rukeya} and \cite{2022Kemp}. The shaded area is the error range of the latter. The results of AGB are represented by purple dotted lines.
\label{fig:4}}
\end{figure}
Using MESA, we construct a grid for each nova eruption with $M_{\rm WD}$, 
$\dot{M}$ and $^7$Li yields.
For every nova simulated using the BSE code, by a bilinear interpolation of the above two physical quantities ($M_{\rm WD}$ and $\dot{M}$) in MESA, one can then calculate the $^7$Li yields of every nova.

Following \cite{2006Lv}, in the method of population synthesis, we use the initial mass function of \cite{1979Miller} for the mass of the primary components and a flat distribution of mass ratios \citep{1989Kraicheva,1994Goldberg}.
A logarithmic flat distribution of initial separations between 10 and 10$^6$ R$_\odot$ is used.

In the binary evolution, the common envelope (CE) evolution is critical important for the formation of nova systems.
In BSE code, a combined parameter $\alpha_{\rm CE}\times\beta_{\rm CE}$ is used to calculate common envelope evolution, where $\alpha_{\rm CE}$ is the fraction of binary binding energy which is spent to expel the CE, and $\beta_{\rm CE}$ is a parameter for the envelope structure of the donor.
As discussed in \cite{2017Rukeya}, $\alpha_{\rm CE}\times\beta_{\rm CE}$ has a weak effect on $^7$Li yields produced by novae.
In this work, we take $\alpha_{\rm CE}\times\beta_{\rm CE}=1.0$.
Simultaneously, circle orbits and solar metallicity for all binary systems are taken.
\textbf{To calculate the production rate of $^7$Li, we assume a constant star formation rate of 5\,M$_{\odot}$\,yr$^{-1}$ over the past 13 Gyr.
In the case of a constant star formation rate, one simulates that $10^{6}$ binary systems, in which the primaries are more massive than 0.8\,$\rm M_{\odot}$, which causes a statistical error for Monte Carlo simulations lower than 1\% in classical novae.}

Figure \ref{fig:4} shows the evolutionary trajectory of $^7$Li produced by novae in the Galaxy at a constant star formation rate.
According to the results of population synthesis, we obtained an average $^7$Li yield of about 6.4\,$\times$\,$10^{-11}$\,$\rm M_{\odot}$ for each nova eruption, with approximately 8,000 eruptions per nova binary, which is consistent with the previous results \citep{1986Shara,2020Molaro}. 
We then select nova systems from binary systems, the primary is a white dwarf (including CO white dwarf and ONeMg white dwarf) and is accreting material from its secondary (a main sequence or red giant star) via Roche lobe or stellar wind.
The result is that among the $10^{6}$ binary systems, 17368 of them evolve into nova systems, accounting for approximately 1.7\%. 
This is consistent with \cite{2017Rukeya} ($\sim$1.5\%) and \cite{2002Hurley} ($\sim$1.9\%).
Based on this, we can conclude that the nova rate is 130 per year, which is within the predicted range \citep{1972Sharov,1994della,1997Shafter,2002Shafter,2017Shafter}.
Among the $10^{6}$ binary systems, 1.7\% of them evolve into nova systems, resulting in an annual $^7$Li production of approximately 8.4\,$\times$\,$10^{-9}$\,$\rm M_{\odot}$.
Considering the age of the Galaxy to be $1.3\times10^{10}$ years, the contribution of novae to $^7$Li is estimated to be around 110\,$\rm M_{\odot}$, compared to the total Li content of approximately 150\,$\rm M_{\odot}$ in the interstellar medium of the Galaxy \citep{1996Hernanz,2016Molaro}. 
Therefore, novae contribute approximately 73\% of the $^7$Li in the Galactic ISM.

In Figure \ref{fig:4}, the solid red line represents the $^7$Li production at the age scale of the Galaxy, which is significantly higher than the results obtained by \cite{2017Rukeya}.
\cite{2022Kemp} used the binary population synthesis code \textit{binary\_c}, to simulate the novae in the Galactic and, by using observed Li production, effectively explained the Li abundance in the early Sun.
The $^7$Li production calculated in our model falls within the predicted error range for nova $^7$Li production, as demonstrated by \cite{2022Kemp}, proving that novae are indeed the most significant contributors to Li in the Galaxy.
In addition, we also calculate the contribution of asymptotic giant branch (AGB) stars using the same method.
The hot bottom burning mechanism of AGB stars can bring Li elements to the surface, which will evolve into the planetary nebulae stage and ultimately be ejected into the ISM \citep{1990Smith,2001Romano}.
After calculation, each AGB star produces approximately $10^{-9}$\,$\rm M_{\odot}$ of $^7$Li, resulting in a final contribution to $^7$Li of around 1\,$\rm M_{\odot}$, which is only 1\%. 
This result has been consistent with previous research \citep{2001Romano,2014Doherty,2017Rukeya}.

In our simulation, the lithium mass produced by novae is $\sim$ 110 M$_\odot$. 
It indicates that the novae can provide about 70\% of the contribution to the Galactic ISM.
However, compared to the total Li mass in the entire Galaxy ($\sim$ 1000 M$_\odot$) \citep{2023Molaro}, Li produced by the novae is limited (about 15\%-20\%), and most of Li may originate from the Big Bang. 

\section{summary}
We used the MESA code to construct nova models with element diffusion to calculate the abundance of $^7$Be on the surface of the WD during the occurrence of a TNR, as well as the production of $^7$Li after the eruption.
Element diffusion enhances the efficiency of transport between the nuclear reaction zone and the convective region on the surface of the WD, allowing more material to be transported to the surface.
Based on theoretical and observational evidence, we appropriately increased the amount of $^3$He produced during the mixing of material from the donor star and the accretor star, leading to more $^7$Be being produced.
A transport channel within the WD is formed due to the increased transmission efficiency caused by element diffusion.
This channel effectively transports the $^7$Be from the hydrogen-burning region of the WD to the convective envelope, where it decays into $^7$Li.
Therefore, a large amount of $^7$Be produced by the nuclear reaction process in the WD can be transferred to the surface and ultimately ejected.
The abundance of $^7$Be on the surface of the WD during the occurrence of a TNR in our model is consistent with the values obtained from the observed nova samples.

When the $^7$Be ejected by the nova eruption decays into $^7$Li in a short period of time, we use population synthesis methods to calculate the production of $^7$Li from the nova eruption. 
We find that about 1.7\% binary systems in the Galaxy can evolve into nova binaries.
Every nova binary evenly can occur about 8,000 eruptions, and the occurrence of nova eruption is about 130 yr$^{-1}$. Each eruption can produce about 6.4\,$\times$\,$10^{-11}$\,M$_{\odot}$ of $^7$Li.
There are about 110 M$_\odot$ Li produced by novae, which is about 73\% of total Li mass in the interstellar medium of the Galaxy, and approximately 15\%-20\% of the entire Galaxy.
It means that novae are the primary source of Li in the Galactic ISM, and also one of the important sources of lithium in the entire Galaxy.

\section*{Acknowledgements}
This work received the generous
support of the National Natural Science Foundation of China
under grants 12163005, U2031204, 12373038 and 12288102, the science
research grants from the China Manned Space Project with No.
CMSCSST-2021-A10 and the Natural Science Foundation of Xinjiang Nos.
2021D01C075, No.2022D01D85, and 2022TSYCLJ0006.


\bibliography{sample631}{}

\begin{thebibliography}{}
\expandafter\ifx\csname natexlab\endcsname\relax\def\natexlab#1{#1}\fi
\providecommand{\url}[1]{\href{#1}{#1}}
\providecommand{\dodoi}[1]{doi:~\href{http://doi.org/#1}{\nolinkurl{#1}}}
\providecommand{\doeprint}[1]{\href{http://ascl.net/#1}{\nolinkurl{http://ascl.net/#1}}}
\providecommand{\doarXiv}[1]{\href{https://arxiv.org/abs/#1}{\nolinkurl{https://arxiv.org/abs/#1}}}

\bibitem[{{Alib{\'e}s} {et~al.}(2002){Alib{\'e}s}, {Labay}, \& {Canal}}]{2002Alib}
{Alib{\'e}s}, A., {Labay}, J., \& {Canal}, R. 2002, \apj, 571, 326, \dodoi{10.1086/339937}

\bibitem[{{Arai} {et~al.}(2021){Arai}, {Tajitsu}, {Kawakita}, \& {Shinnaka}}]{2021Arai}
{Arai}, A., {Tajitsu}, A., {Kawakita}, H., \& {Shinnaka}, Y. 2021, \apj, 916, 44, \dodoi{10.3847/1538-4357/ac00bf}

\bibitem[{{Arnould} \& {Norgaard}(1975)}]{1975Arnould}
{Arnould}, M., \& {Norgaard}, H. 1975, \aap, 42, 55

\bibitem[{{Asplund} {et~al.}(2009){Asplund}, {Grevesse}, {Sauval}, \& {Scott}}]{2009Asplund}
{Asplund}, M., {Grevesse}, N., {Sauval}, A.~J., \& {Scott}, P. 2009, \araa, 47, 481, \dodoi{10.1146/annurev.astro.46.060407.145222}

\bibitem[{{Balser} \& {Bania}(2018)}]{2018Balser}
{Balser}, D.~S., \& {Bania}, T.~M. 2018, \aj, 156, 280, \dodoi{10.3847/1538-3881/aaeb2b}

\bibitem[{{Banerjee, Projjwal} {et~al.}(2016){Banerjee, Projjwal}, {Qian, Yong-Zhong}, {Heger, Alexander}, \& {Haxton, Wick}}]{2016Banerjee}
{Banerjee, Projjwal}, {Qian, Yong-Zhong}, {Heger, Alexander}, \& {Haxton, Wick}. 2016, EPJ Web of Conferences, 109, 06001, \dodoi{10.1051/epjconf/201610906001}

\bibitem[{{Bennett} {et~al.}(2013){Bennett}, {Larson}, {Weiland}, {Jarosik}, {Hinshaw}, {Odegard}, {Smith}, {Hill}, {Gold}, {Halpern}, {Komatsu}, {Nolta}, {Page}, {Spergel}, {Wollack}, {Dunkley}, {Kogut}, {Limon}, {Meyer}, {Tucker}, \& {Wright}}]{2013WMAP}
{Bennett}, C.~L., {Larson}, D., {Weiland}, J.~L., {et~al.} 2013, \apjs, 208, 20, \dodoi{10.1088/0067-0049/208/2/20}

\bibitem[{{Boffin} {et~al.}(1993){Boffin}, {Paulus}, {Arnould}, \& {Mowlavi}}]{1993Boffin}
{Boffin}, H.~M.~J., {Paulus}, G., {Arnould}, M., \& {Mowlavi}, N. 1993, \aap, 279, 173

\bibitem[{{Burgers}(1969)}]{1969Burgers}
{Burgers}, J.~M. 1969, {Flow Equations for Composite Gases}

\bibitem[{{Cameron} \& {Fowler}(1971)}]{1971Cameron}
{Cameron}, A.~G.~W., \& {Fowler}, W.~A. 1971, \apj, 164, 111, \dodoi{10.1086/150821}

\bibitem[{{Casey} {et~al.}(2016){Casey}, {Ruchti}, {Masseron}, {Randich}, {Gilmore}, {Lind}, {Kennedy}, {Koposov}, {Hourihane}, {Franciosini}, {Lewis}, {Magrini}, {Morbidelli}, {Sacco}, {Worley}, {Feltzing}, {Jeffries}, {Vallenari}, {Bensby}, {Bragaglia}, {Flaccomio}, {Francois}, {Korn}, {Lanzafame}, {Pancino}, {Recio-Blanco}, {Smiljanic}, {Carraro}, {Costado}, {Damiani}, {Donati}, {Frasca}, {Jofr{\'e}}, {Lardo}, {de Laverny}, {Monaco}, {Prisinzano}, {Sbordone}, {Sousa}, {Tautvai{\v{s}}ien{\.{e}}}, {Zaggia}, {Zwitter}, {Delgado Mena}, {Chorniy}, {Martell}, {Silva Aguirre}, {Miglio}, {Chiappini}, {Montalban}, {Morel}, \& {Valentini}}]{2016Casey}
{Casey}, A.~R., {Ruchti}, G., {Masseron}, T., {et~al.} 2016, \mnras, 461, 3336, \dodoi{10.1093/mnras/stw1512}

\bibitem[{{Cescutti} \& {Molaro}(2019)}]{2019Cescutti}
{Cescutti}, G., \& {Molaro}, P. 2019, \mnras, 482, 4372, \dodoi{10.1093/mnras/sty2967}

\bibitem[{{Coc} {et~al.}(2014){Coc}, {Uzan}, \& {Vangioni}}]{2014Coc}
{Coc}, A., {Uzan}, J.-P., \& {Vangioni}, E. 2014, \jcap, 2014, 050, \dodoi{10.1088/1475-7516/2014/10/050}

\bibitem[{{Dantona} \& {Mazzitelli}(1982)}]{1982Dantona}
{Dantona}, F., \& {Mazzitelli}, I. 1982, \apj, 260, 722, \dodoi{10.1086/160292}

\bibitem[{{Dearborn} {et~al.}(1996){Dearborn}, {Steigman}, \& {Tosi}}]{1996Dearborn}
{Dearborn}, D. S.~P., {Steigman}, G., \& {Tosi}, M. 1996, \apj, 465, 887, \dodoi{10.1086/177472}

\bibitem[{{della Valle} \& {Livio}(1994)}]{1994della}
{della Valle}, M., \& {Livio}, M. 1994, \aap, 286, 786

\bibitem[{{Denissenkov} {et~al.}(2013){Denissenkov}, {Herwig}, {Bildsten}, \& {Paxton}}]{2013Denissenkov}
{Denissenkov}, P.~A., {Herwig}, F., {Bildsten}, L., \& {Paxton}, B. 2013, \apj, 762, 8, \dodoi{10.1088/0004-637X/762/1/8}

\bibitem[{{Denissenkov} {et~al.}(2021){Denissenkov}, {Ruiz}, {Upadhyayula}, \& {Herwig}}]{2021Denissenkov}
{Denissenkov}, P.~A., {Ruiz}, C., {Upadhyayula}, S., \& {Herwig}, F. 2021, \mnras, 501, L33, \dodoi{10.1093/mnrasl/slaa190}

\bibitem[{{Denissenkov} {et~al.}(2014){Denissenkov}, {Truran}, {Pignatari}, {Trappitsch}, {Ritter}, {Herwig}, {Battino}, {Setoodehnia}, \& {Paxton}}]{2014Denissenkov}
{Denissenkov}, P.~A., {Truran}, J.~W., {Pignatari}, M., {et~al.} 2014, \mnras, 442, 2058, \dodoi{10.1093/mnras/stu1000}

\bibitem[{{Doherty} {et~al.}(2014){Doherty}, {Gil-Pons}, {Lau}, {Lattanzio}, {Siess}, \& {Campbell}}]{2014Doherty}
{Doherty}, C.~L., {Gil-Pons}, P., {Lau}, H. H.~B., {et~al.} 2014, \mnras, 441, 582, \dodoi{10.1093/mnras/stu571}

\bibitem[{{Draine}(2011)}]{2011Draine}
{Draine}, B.~T. 2011, {Physics of the Interstellar and Intergalactic Medium}

\bibitem[{{Dupuis} {et~al.}(1992){Dupuis}, {Fontaine}, {Pelletier}, \& {Wesemael}}]{1992Dupuis}
{Dupuis}, J., {Fontaine}, G., {Pelletier}, C., \& {Wesemael}, F. 1992, \apjs, 82, 505, \dodoi{10.1086/191728}

\bibitem[{{Fields} {et~al.}(2014){Fields}, {Molaro}, \& {Sarkar}}]{2014Fields}
{Fields}, B.~D., {Molaro}, P., \& {Sarkar}, S. 2014, arXiv e-prints, arXiv:1412.1408, \dodoi{10.48550/arXiv.1412.1408}

\bibitem[{{Gao} {et~al.}(2022){Gao}, {Zhu}, {Yu}, {Liu}, {Lu}, {Shi}, \& {L{\"u}}}]{2022Gao}
{Gao}, J., {Zhu}, C., {Yu}, J., {et~al.} 2022, \aap, 668, A126, \dodoi{10.1051/0004-6361/202243871}

\bibitem[{{Gehrz} {et~al.}(1998){Gehrz}, {Truran}, {Williams}, \& {Starrfield}}]{1998Gehrz}
{Gehrz}, R.~D., {Truran}, J.~W., {Williams}, R.~E., \& {Starrfield}, S. 1998, \pasp, 110, 3, \dodoi{10.1086/316107}

\bibitem[{{Goldberg} \& {Mazeh}(1994)}]{1994Goldberg}
{Goldberg}, D., \& {Mazeh}, T. 1994, \aap, 282, 801

\bibitem[{{Han} {et~al.}(2020){Han}, {Ge}, {Chen}, \& {Chen}}]{2020Han}
{Han}, Z.-W., {Ge}, H.-W., {Chen}, X.-F., \& {Chen}, H.-L. 2020, Research in Astronomy and Astrophysics, 20, 161, \dodoi{10.1088/1674-4527/20/10/161}

\bibitem[{{Hernanz} {et~al.}(1996){Hernanz}, {Jose}, {Coc}, \& {Isern}}]{1996Hernanz}
{Hernanz}, M., {Jose}, J., {Coc}, A., \& {Isern}, J. 1996, \apjl, 465, L27, \dodoi{10.1086/310122}

\bibitem[{{Hurley} {et~al.}(2002){Hurley}, {Tout}, \& {Pols}}]{2002Hurley}
{Hurley}, J.~R., {Tout}, C.~A., \& {Pols}, O.~R. 2002, \mnras, 329, 897, \dodoi{10.1046/j.1365-8711.2002.05038.x}

\bibitem[{{Iben} \& {Tutukov}(1984)}]{1984Iben}
{Iben}, I., J., \& {Tutukov}, A.~V. 1984, \apj, 284, 719, \dodoi{10.1086/162455}

\bibitem[{{Iben} {et~al.}(1991){Iben}, {Fujimoto}, \& {MacDonald}}]{1991Iben}
{Iben}, Icko, J., {Fujimoto}, M.~Y., \& {MacDonald}, J. 1991, \apjl, 375, L27, \dodoi{10.1086/186080}

\bibitem[{{Izzo} {et~al.}(2015){Izzo}, {Della Valle}, {Mason}, {Matteucci}, {Romano}, {Pasquini}, {Vanzi}, {Jordan}, {Fernandez}, {Bluhm}, {Brahm}, {Espinoza}, \& {Williams}}]{2015Izzo}
{Izzo}, L., {Della Valle}, M., {Mason}, E., {et~al.} 2015, \apjl, 808, L14, \dodoi{10.1088/2041-8205/808/1/L14}

\bibitem[{{Izzo} {et~al.}(2018){Izzo}, {Molaro}, {Bonifacio}, {Della Valle}, {Cano}, {de Ugarte Postigo}, {Prieto}, {Th{\"o}ne}, {Vanzi}, {Zapata}, \& {Fernandez}}]{2018Izzo}
{Izzo}, L., {Molaro}, P., {Bonifacio}, P., {et~al.} 2018, \mnras, 478, 1601, \dodoi{10.1093/mnras/sty435}

\bibitem[{{Jose}(2016)}]{2016Jose}
{Jose}, J. 2016, {Stellar Explosions: Hydrodynamics and Nucleosynthesis}, \dodoi{10.1201/b19165}

\bibitem[{{Jos{\'e}} \& {Hernanz}(1998)}]{1998Jos}
{Jos{\'e}}, J., \& {Hernanz}, M. 1998, \apj, 494, 680, \dodoi{10.1086/305244}

\bibitem[{{Jos{\'e}} {et~al.}(2020){Jos{\'e}}, {Shore}, \& {Casanova}}]{2020Jos}
{Jos{\'e}}, J., {Shore}, S.~N., \& {Casanova}, J. 2020, \aap, 634, A5, \dodoi{10.1051/0004-6361/201936893}

\bibitem[{{Kemp} {et~al.}(2022{\natexlab{a}}){Kemp}, {Karakas}, {Casey}, {C{\^o}t{\'e}}, {Izzard}, \& {Osborn}}]{2022Kemp}
{Kemp}, A.~J., {Karakas}, A.~I., {Casey}, A.~R., {et~al.} 2022{\natexlab{a}}, \apjl, 933, L30, \dodoi{10.3847/2041-8213/ac7c72}

\bibitem[{{Kemp} {et~al.}(2022{\natexlab{b}}){Kemp}, {Karakas}, {Casey}, {Kobayashi}, \& {Izzard}}]{2022Kemp1}
{Kemp}, A.~J., {Karakas}, A.~I., {Casey}, A.~R., {Kobayashi}, C., \& {Izzard}, R.~G. 2022{\natexlab{b}}, \mnras, 509, 1175, \dodoi{10.1093/mnras/stab3103}

\bibitem[{{Kippenhahn} {et~al.}(1980){Kippenhahn}, {Ruschenplatt}, \& {Thomas}}]{1980Kippenhahn}
{Kippenhahn}, R., {Ruschenplatt}, G., \& {Thomas}, H.~C. 1980, \aap, 91, 175

\bibitem[{{Kovetz} \& {Prialnik}(1985)}]{1985Kovetz}
{Kovetz}, A., \& {Prialnik}, D. 1985, \apj, 291, 812, \dodoi{10.1086/163117}

\bibitem[{{Kraicheva} {et~al.}(1989){Kraicheva}, {Popova}, {Tutukov}, \& {Yungel'Son}}]{1989Kraicheva}
{Kraicheva}, Z.~T., {Popova}, E.~I., {Tutukov}, A.~V., \& {Yungel'Son}, L.~R. 1989, Astrophysics, 30, 323, \dodoi{10.1007/BF01003893}

\bibitem[{{Lodders} {et~al.}(2009){Lodders}, {Palme}, \& {Gail}}]{2009Lodders}
{Lodders}, K., {Palme}, H., \& {Gail}, H.~P. 2009, Landolt B\&ouml;rnstein, 4B, 712, \dodoi{10.1007/978-3-540-88055-4_34}

\bibitem[{{L{\"u}} {et~al.}(2006){L{\"u}}, {Yungelson}, \& {Han}}]{2006Lv}
{L{\"u}}, G., {Yungelson}, L., \& {Han}, Z. 2006, \mnras, 372, 1389, \dodoi{10.1111/j.1365-2966.2006.10947.x}

\bibitem[{{L{\"u}} {et~al.}(2013){L{\"u}}, {Zhu}, \& {Podsiadlowski}}]{2013Lv}
{L{\"u}}, G., {Zhu}, C., \& {Podsiadlowski}, P. 2013, \apj, 768, 193, \dodoi{10.1088/0004-637X/768/2/193}

\bibitem[{{L{\"u}} {et~al.}(2020){L{\"u}}, {Zhu}, {Wang}, {Liu}, {Li}, {Xie}, \& {Liu}}]{2020Lv}
{L{\"u}}, G., {Zhu}, C., {Wang}, Z., {et~al.} 2020, \apj, 890, 69, \dodoi{10.3847/1538-4357/ab6bcc}

\bibitem[{{L{\"u}} {et~al.}(2009){L{\"u}}, {Zhu}, {Wang}, \& {Wang}}]{2009Lv}
{L{\"u}}, G., {Zhu}, C., {Wang}, Z., \& {Wang}, N. 2009, \mnras, 396, 1086, \dodoi{10.1111/j.1365-2966.2009.14777.x}

\bibitem[{{Miller} \& {Scalo}(1979)}]{1979Miller}
{Miller}, G.~E., \& {Scalo}, J.~M. 1979, \apjs, 41, 513, \dodoi{10.1086/190629}

\bibitem[{{Molaro} {et~al.}(2020){Molaro}, {Izzo}, {Bonifacio}, {Hernanz}, {Selvelli}, \& {della Valle}}]{2020Molaro}
{Molaro}, P., {Izzo}, L., {Bonifacio}, P., {et~al.} 2020, \mnras, 492, 4975, \dodoi{10.1093/mnras/stz3587}

\bibitem[{{Molaro} {et~al.}(2016){Molaro}, {Izzo}, {Mason}, {Bonifacio}, \& {Della Valle}}]{2016Molaro}
{Molaro}, P., {Izzo}, L., {Mason}, E., {Bonifacio}, P., \& {Della Valle}, M. 2016, \mnras, 463, L117, \dodoi{10.1093/mnrasl/slw169}

\bibitem[{{Molaro} {et~al.}(2022){Molaro}, {Izzo}, {D'Odorico}, {Aydi}, {Bonifacio}, {Cescutti}, {Harvey}, {Hernanz}, {Selvelli}, \& {della Valle}}]{2022Molaro}
{Molaro}, P., {Izzo}, L., {D'Odorico}, V., {et~al.} 2022, \mnras, 509, 3258, \dodoi{10.1093/mnras/stab3106}

\bibitem[{{Molaro} {et~al.}(2023){Molaro}, {Izzo}, {Selvelli}, {Bonifacio}, {Aydi}, {Cescutti}, {Guido}, {Harvey}, {Hernanz}, \& {Della Valle}}]{2023Molaro}
{Molaro}, P., {Izzo}, L., {Selvelli}, P., {et~al.} 2023, \mnras, 518, 2614, \dodoi{10.1093/mnras/stac2708}

\bibitem[{Mukai \& Sokoloski(2019)}]{2019Mukai}
Mukai, K., \& Sokoloski, J.~L. 2019, Physics Today, 72, 38, \dodoi{10.1063/PT.3.4341}

\bibitem[{{Paxton} {et~al.}(2011){Paxton}, {Bildsten}, {Dotter}, {Herwig}, {Lesaffre}, \& {Timmes}}]{2011Paxton}
{Paxton}, B., {Bildsten}, L., {Dotter}, A., {et~al.} 2011, \apjs, 192, 3, \dodoi{10.1088/0067-0049/192/1/3}

\bibitem[{{Paxton} {et~al.}(2013){Paxton}, {Cantiello}, {Arras}, {Bildsten}, {Brown}, {Dotter}, {Mankovich}, {Montgomery}, {Stello}, {Timmes}, \& {Townsend}}]{2013Paxton}
{Paxton}, B., {Cantiello}, M., {Arras}, P., {et~al.} 2013, \apjs, 208, 4, \dodoi{10.1088/0067-0049/208/1/4}

\bibitem[{{Paxton} {et~al.}(2015){Paxton}, {Marchant}, {Schwab}, {Bauer}, {Bildsten}, {Cantiello}, {Dessart}, {Farmer}, {Hu}, {Langer}, {Townsend}, {Townsley}, \& {Timmes}}]{2015Paxton}
{Paxton}, B., {Marchant}, P., {Schwab}, J., {et~al.} 2015, \apjs, 220, 15, \dodoi{10.1088/0067-0049/220/1/15}

\bibitem[{{Paxton} {et~al.}(2018){Paxton}, {Schwab}, {Bauer}, {Bildsten}, {Blinnikov}, {Duffell}, {Farmer}, {Goldberg}, {Marchant}, {Sorokina}, {Thoul}, {Townsend}, \& {Timmes}}]{2018Paxton}
{Paxton}, B., {Schwab}, J., {Bauer}, E.~B., {et~al.} 2018, \apjs, 234, 34, \dodoi{10.3847/1538-4365/aaa5a8}

\bibitem[{{Paxton} {et~al.}(2019){Paxton}, {Smolec}, {Schwab}, {Gautschy}, {Bildsten}, {Cantiello}, {Dotter}, {Farmer}, {Goldberg}, {Jermyn}, {Kanbur}, {Marchant}, {Thoul}, {Townsend}, {Wolf}, {Zhang}, \& {Timmes}}]{2019Paxton}
{Paxton}, B., {Smolec}, R., {Schwab}, J., {et~al.} 2019, \apjs, 243, 10, \dodoi{10.3847/1538-4365/ab2241}

\bibitem[{{Pignatari} {et~al.}(2016){Pignatari}, {Herwig}, {Hirschi}, {Bennett}, {Rockefeller}, {Fryer}, {Timmes}, {Ritter}, {Heger}, {Jones}, {Battino}, {Dotter}, {Trappitsch}, {Diehl}, {Frischknecht}, {Hungerford}, {Magkotsios}, {Travaglio}, \& {Young}}]{2016Pignatari}
{Pignatari}, M., {Herwig}, F., {Hirschi}, R., {et~al.} 2016, \apjs, 225, 24, \dodoi{10.3847/0067-0049/225/2/24}

\bibitem[{{Planck Collaboration} {et~al.}(2020){Planck Collaboration}, {Aghanim}, {Akrami}, {Ashdown}, {Aumont}, {Baccigalupi}, {Ballardini}, {Banday}, {Barreiro}, {Bartolo}, {Basak}, {Battye}, {Benabed}, {Bernard}, {Bersanelli}, {Bielewicz}, {Bock}, {Bond}, {Borrill}, {Bouchet}, {Boulanger}, {Bucher}, {Burigana}, {Butler}, {Calabrese}, {Cardoso}, {Carron}, {Challinor}, {Chiang}, {Chluba}, {Colombo}, {Combet}, {Contreras}, {Crill}, {Cuttaia}, {de Bernardis}, {de Zotti}, {Delabrouille}, {Delouis}, {Di Valentino}, {Diego}, {Dor{\'e}}, {Douspis}, {Ducout}, {Dupac}, {Dusini}, {Efstathiou}, {Elsner}, {En{\ss}lin}, {Eriksen}, {Fantaye}, {Farhang}, {Fergusson}, {Fernandez-Cobos}, {Finelli}, {Forastieri}, {Frailis}, {Fraisse}, {Franceschi}, {Frolov}, {Galeotta}, {Galli}, {Ganga}, {G{\'e}nova-Santos}, {Gerbino}, {Ghosh}, {Gonz{\'a}lez-Nuevo}, {G{\'o}rski}, {Gratton}, {Gruppuso}, {Gudmundsson}, {Hamann}, {Handley}, {Hansen}, {Herranz}, {Hildebrandt}, {Hivon}, {Huang}, {Jaffe}, {Jones}, {Karakci}, {Keih{\"a}nen},
  {Keskitalo}, {Kiiveri}, {Kim}, {Kisner}, {Knox}, {Krachmalnicoff}, {Kunz}, {Kurki-Suonio}, {Lagache}, {Lamarre}, {Lasenby}, {Lattanzi}, {Lawrence}, {Le Jeune}, {Lemos}, {Lesgourgues}, {Levrier}, {Lewis}, {Liguori}, {Lilje}, {Lilley}, {Lindholm}, {L{\'o}pez-Caniego}, {Lubin}, {Ma}, {Mac{\'\i}as-P{\'e}rez}, {Maggio}, {Maino}, {Mandolesi}, {Mangilli}, {Marcos-Caballero}, {Maris}, {Martin}, {Martinelli}, {Mart{\'\i}nez-Gonz{\'a}lez}, {Matarrese}, {Mauri}, {McEwen}, {Meinhold}, {Melchiorri}, {Mennella}, {Migliaccio}, {Millea}, {Mitra}, {Miville-Desch{\^e}nes}, {Molinari}, {Montier}, {Morgante}, {Moss}, {Natoli}, {N{\o}rgaard-Nielsen}, {Pagano}, {Paoletti}, {Partridge}, {Patanchon}, {Peiris}, {Perrotta}, {Pettorino}, {Piacentini}, {Polastri}, {Polenta}, {Puget}, {Rachen}, {Reinecke}, {Remazeilles}, {Renzi}, {Rocha}, {Rosset}, {Roudier}, {Rubi{\~n}o-Mart{\'\i}n}, {Ruiz-Granados}, {Salvati}, {Sandri}, {Savelainen}, {Scott}, {Shellard}, {Sirignano}, {Sirri}, {Spencer}, {Sunyaev}, {Suur-Uski}, {Tauber}, {Tavagnacco},
  {Tenti}, {Toffolatti}, {Tomasi}, {Trombetti}, {Valenziano}, {Valiviita}, {Van Tent}, {Vibert}, {Vielva}, {Villa}, {Vittorio}, {Wandelt}, {Wehus}, {White}, {White}, {Zacchei}, \& {Zonca}}]{2020Planck}
{Planck Collaboration}, {Aghanim}, N., {Akrami}, Y., {et~al.} 2020, \aap, 641, A6, \dodoi{10.1051/0004-6361/201833910}

\bibitem[{{Prantzos}(2012)}]{2012Prantzos}
{Prantzos}, N. 2012, \aap, 542, A67, \dodoi{10.1051/0004-6361/201219043}

\bibitem[{{Prantzos} {et~al.}(1993){Prantzos}, {Casse}, \& {Vangioni-Flam}}]{1993Prantzos}
{Prantzos}, N., {Casse}, M., \& {Vangioni-Flam}, E. 1993, \apj, 403, 630, \dodoi{10.1086/172233}

\bibitem[{{Prantzos, N.} {et~al.}(2017){Prantzos, N.}, {de Laverny, P.}, {Guiglion, G.}, {Recio-Blanco, A.}, \& {Worley, C. C.}}]{2017Prantzos}
{Prantzos, N.}, {de Laverny, P.}, {Guiglion, G.}, {Recio-Blanco, A.}, \& {Worley, C. C.} 2017, A\&A, 606, A132, \dodoi{10.1051/0004-6361/201731188}

\bibitem[{{Romano} \& {Matteucci}(2003)}]{2003Romano}
{Romano}, D., \& {Matteucci}, F. 2003, \mnras, 342, 185, \dodoi{10.1046/j.1365-8711.2003.06526.x}

\bibitem[{{Romano} {et~al.}(1999){Romano}, {Matteucci}, {Molaro}, \& {Bonifacio}}]{1999Romano}
{Romano}, D., {Matteucci}, F., {Molaro}, P., \& {Bonifacio}, P. 1999, \aap, 352, 117, \dodoi{10.48550/arXiv.astro-ph/9910151}

\bibitem[{{Romano} {et~al.}(2001){Romano}, {Matteucci}, {Ventura}, \& {D'Antona}}]{2001Romano}
{Romano}, D., {Matteucci}, F., {Ventura}, P., \& {D'Antona}, F. 2001, \aap, 374, 646, \dodoi{10.1051/0004-6361:20010751}

\bibitem[{{Rood} {et~al.}(1992){Rood}, {Bania}, \& {Wilson}}]{1992Rood}
{Rood}, R.~T., {Bania}, T.~M., \& {Wilson}, T.~L. 1992, \nat, 355, 618, \dodoi{10.1038/355618a0}

\bibitem[{{Rukeya} {et~al.}(2017){Rukeya}, {L{\"u}}, {Wang}, \& {Zhu}}]{2017Rukeya}
{Rukeya}, R., {L{\"u}}, G., {Wang}, Z., \& {Zhu}, C. 2017, \pasp, 129, 074201, \dodoi{10.1088/1538-3873/aa6b4d}

\bibitem[{{Schatzman}(1951)}]{1951Schatzman}
{Schatzman}, E. 1951, Annales d'Astrophysique, 14, 294

\bibitem[{{Selvelli} {et~al.}(2018){Selvelli}, {Molaro}, \& {Izzo}}]{2018Selvelli}
{Selvelli}, P., {Molaro}, P., \& {Izzo}, L. 2018, \mnras, 481, 2261, \dodoi{10.1093/mnras/sty2310}

\bibitem[{{Shafter}(2017)}]{2017Shafter}
{Shafter}. 2017, \apj, 834, 196, \dodoi{10.3847/1538-4357/834/2/196}

\bibitem[{{Shafter}(1997)}]{1997Shafter}
{Shafter}, A.~W. 1997, \apj, 487, 226, \dodoi{10.1086/304609}

\bibitem[{{Shafter}(2002)}]{2002Shafter}
{Shafter}, A.~W. 2002, in American Institute of Physics Conference Series, Vol. 637, Classical Nova Explosions, ed. M.~{Hernanz} \& J.~{Jos{\'e}}, 462--471, \dodoi{10.1063/1.1518246}

\bibitem[{{Shara}(1980)}]{1980Shara}
{Shara}, M.~M. 1980, \apj, 239, 581, \dodoi{10.1086/158144}

\bibitem[{{Shara} {et~al.}(1986){Shara}, {Livio}, {Moffat}, \& {Orio}}]{1986Shara}
{Shara}, M.~M., {Livio}, M., {Moffat}, A. F.~J., \& {Orio}, M. 1986, \apj, 311, 163, \dodoi{10.1086/164762}

\bibitem[{{Sharov}(1972)}]{1972Sharov}
{Sharov}, A.~S. 1972, \sovast, 16, 41

\bibitem[{{Shen} \& {Bildsten}(2009)}]{2009Shen}
{Shen}, K.~J., \& {Bildsten}, L. 2009, \apj, 692, 324, \dodoi{10.1088/0004-637X/692/1/324}

\bibitem[{{Smith} \& {Lambert}(1989)}]{1989Smith}
{Smith}, V.~V., \& {Lambert}, D.~L. 1989, \apjl, 345, L75, \dodoi{10.1086/185556}

\bibitem[{{{Smith}, Verne V. and {Lambert}, David L.}(1990)}]{1990Smith}
{{Smith}, Verne V. and {Lambert}, David L.} 1990, \apjs, 72, 387, \dodoi{10.1086/191421}

\bibitem[{{Spergel} {et~al.}(2003){Spergel}, {Verde}, {Peiris}, {Komatsu}, {Nolta}, {Bennett}, {Halpern}, {Hinshaw}, {Jarosik}, {Kogut}, {Limon}, {Meyer}, {Page}, {Tucker}, {Weiland}, {Wollack}, \& {Wright}}]{2003WMAP}
{Spergel}, D.~N., {Verde}, L., {Peiris}, H.~V., {et~al.} 2003, \apjs, 148, 175, \dodoi{10.1086/377226}

\bibitem[{{Spite} \& {Spite}(1982)}]{1982Spite}
{Spite}, M., \& {Spite}, F. 1982, \nat, 297, 483, \dodoi{10.1038/297483a0}

\bibitem[{{Starrfield} {et~al.}(2020){Starrfield}, {Bose}, {Iliadis}, {Hix}, {Woodward}, \& {Wagner}}]{2020Starrfield}
{Starrfield}, S., {Bose}, M., {Iliadis}, C., {et~al.} 2020, \apj, 895, 70, \dodoi{10.3847/1538-4357/ab8d23}

\bibitem[{Starrfield {et~al.}(2024)Starrfield, Bose, Iliadis, Hix, Woodward, \& Wagner}]{starrfield2024}
Starrfield, S., Bose, M., Iliadis, C., {et~al.} 2024, Hydrodynamic Simulations of Oxygen-Neon Classical Novae as Galactic $^7$Li Producers and Potential Accretion Induced Collapse Progenitors.
\newblock \doarXiv{2401.02307}

\bibitem[{{Starrfield} {et~al.}(2016){Starrfield}, {Iliadis}, \& {Hix}}]{2016Starrfield}
{Starrfield}, S., {Iliadis}, C., \& {Hix}, W.~R. 2016, \pasp, 128, 051001, \dodoi{10.1088/1538-3873/128/963/051001}

\bibitem[{{Starrfield} {et~al.}(2009){Starrfield}, {Iliadis}, {Hix}, {Timmes}, \& {Sparks}}]{2009Starrfield}
{Starrfield}, S., {Iliadis}, C., {Hix}, W.~R., {Timmes}, F.~X., \& {Sparks}, W.~M. 2009, \apj, 692, 1532, \dodoi{10.1088/0004-637X/692/2/1532}

\bibitem[{{Starrfield} {et~al.}(1978){Starrfield}, {Truran}, {Sparks}, \& {Arnould}}]{1978Starrfield}
{Starrfield}, S., {Truran}, J.~W., {Sparks}, W.~M., \& {Arnould}, M. 1978, \apj, 222, 600, \dodoi{10.1086/156175}

\bibitem[{{Starrfield} {et~al.}(1998){Starrfield}, {Truran}, {Wiescher}, \& {Sparks}}]{1998Starrfield}
{Starrfield}, S., {Truran}, J.~W., {Wiescher}, M.~C., \& {Sparks}, W.~M. 1998, \mnras, 296, 502, \dodoi{10.1046/j.1365-8711.1998.01312.x}

\bibitem[{{Tajitsu} {et~al.}(2015){Tajitsu}, {Sadakane}, {Naito}, {Arai}, \& {Aoki}}]{2015Tajitsu}
{Tajitsu}, A., {Sadakane}, K., {Naito}, H., {Arai}, A., \& {Aoki}, W. 2015, \nat, 518, 381, \dodoi{10.1038/nature14161}

\bibitem[{{Tajitsu} {et~al.}(2016){Tajitsu}, {Sadakane}, {Naito}, {Arai}, {Kawakita}, \& {Aoki}}]{2016Tajitsu}
{Tajitsu}, A., {Sadakane}, K., {Naito}, H., {et~al.} 2016, \apj, 818, 191, \dodoi{10.3847/0004-637X/818/2/191}

\bibitem[{{Thoul} {et~al.}(1994){Thoul}, {Bahcall}, \& {Loeb}}]{1994Thoul}
{Thoul}, A.~A., {Bahcall}, J.~N., \& {Loeb}, A. 1994, \apj, 421, 828, \dodoi{10.1086/173695}

\bibitem[{{Townsley} \& {Bildsten}(2004)}]{2004Townsley}
{Townsley}, D.~M., \& {Bildsten}, L. 2004, \apj, 600, 390, \dodoi{10.1086/379647}

\bibitem[{{Travaglio} {et~al.}(2001){Travaglio}, {Randich}, {Galli}, {Lattanzio}, {Elliott}, {Forestini}, \& {Ferrini}}]{2001Travaglio}
{Travaglio}, C., {Randich}, S., {Galli}, D., {et~al.} 2001, \apj, 559, 909, \dodoi{10.1086/322415}

\bibitem[{{Yan} {et~al.}(2018){Yan}, {Shi}, {Zhou}, {Chen}, {Li}, {Zhang}, {Bi}, {Wu}, {Li}, {Guo}, {Liu}, {Gao}, {Zhang}, {Zhou}, {Li}, \& {Zhao}}]{2018Yan}
{Yan}, H.-L., {Shi}, J.-R., {Zhou}, Y.-T., {et~al.} 2018, Nature Astronomy, 2, 790, \dodoi{10.1038/s41550-018-0544-7}

\bibitem[{{Yaron} {et~al.}(2005){Yaron}, {Prialnik}, {Shara}, \& {Kovetz}}]{2005Yaron}
{Yaron}, O., {Prialnik}, D., {Shara}, M.~M., \& {Kovetz}, A. 2005, \apj, 623, 398, \dodoi{10.1086/428435}

\bibitem[{{Yu} {et~al.}(2021){Yu}, {Zhang}, \& {L{\"u}}}]{2021Yu}
{Yu}, J., {Zhang}, X., \& {L{\"u}}, G. 2021, \mnras, 504, 2670, \dodoi{10.1093/mnras/stab1063}

\bibitem[{{Yu} {et~al.}(2019){Yu}, {Li}, {Zhu}, {Wang}, {Liu}, {Guo}, {Han}, {Chen}, \& {L{\"u}}}]{2019Yu}
{Yu}, J., {Li}, Z., {Zhu}, C., {et~al.} 2019, \apj, 885, 20, \dodoi{10.3847/1538-4357/ab44b5}

\bibitem[{{Zhu} {et~al.}(2021){Zhu}, {Liu}, {Wang}, \& {L{\"u}}}]{2021Zhu}
{Zhu}, C., {Liu}, H., {Wang}, Z., \& {L{\"u}}, G. 2021, \aap, 654, A57, \dodoi{10.1051/0004-6361/202039692}

\bibitem[{{{\.Z}yczkowski} {et~al.}(1998){{\.Z}yczkowski}, {Horodecki}, {Sanpera}, \& {Lewenstein}}]{1998Zyczkowski}
{{\.Z}yczkowski}, K., {Horodecki}, P., {Sanpera}, A., \& {Lewenstein}, M. 1998, \pra, 58, 883, \dodoi{10.1103/PhysRevA.58.883}

\end{thebibliography}
\bibliographystyle{aasjournal}



\end{document}